\documentclass{iaus}
\usepackage{graphicx}

\title[Masers in the Nuclear Disk] %% give here short title %%
{Masers as Probes of Massive Star Formation in the Nuclear Disk}

\author[Yusef-Zadeh et al.]   %% give here short author list %%
{F. Yusef-Zadeh$^1$,
R. G. Arendt$^2$, C. O. Heinke$^1$, J. L. Hinz$^3$, 
J. W. Hewitt$^1$, 
P. Pratap$^4$, S. V. Ramirez$^5$, 
G. H. Rieke$^3$, D. A. Roberts$^1$, 
S. R. Stolovy$^5$, M. Wardle$^6$ and B. A. Whitney$^7$}

\affiliation{$^1$Department of Physics and Astronomy, Northwestern University, Evanston, 
IL 60208  USA \break email: zadeh@northwestern.edu, j-hewitt@northwestern.edu, 
cheinke@northwestern.edu, doug-roberts@northwestern.edu\\[\affilskip]
$^2$CRESST/UMBC/GSFC, Code 665, Greenbelt, MD 20771 USA
\break email: arendt@milkyway.gsfc.nasa.gov\\
$^3$Steward Observatory, University of Arizona, 933 N. Cherry Ave., Tucson, AZ 85721   USA 
\break email: grieke@as.arizona.edu, jhinz@as.arizona.edu \\
$^4$MIT Haystack Observatory, Westford, MA 01886
 USA \break email: ppratap@haystack.mit.edu \\
$^5$IPAC, Cal Tech, Pasadena, CA 91125  USA \break email: 
solange@ipac.caltech.edu, stolovy@ipac.caltech.edu\\
$^6$Department of Physics, Macquarie University, Sydney NSW 2109,
Australia  \break email: wardle@ics.mq.edu.au\\
$^7$Space Science Institute, 4750 Walnut Street, Suite 205, Boulder, CO 80301 USA \break 
email: bwhitney@spacescience.org\\}

\pubyear{2007}
\volume{242}  %% insert here IAU Symposium No.
\pagerange{1--8}
\date{April 25, 2007 and in revised form ??}
\jname{ Astrophysical Masers and Their Environments}
\editors{J.  Chapman \& W. Baan, eds.}
\begin{document}
\maketitle
\def\kms {\hbox{km{\hskip0.1em}s$^{-1}$}} % km/s
\def\msol{\hbox{$\hbox{M}_\odot$}}
\def\lsol{\hbox{$\hbox{L}_\odot$}}
\def\kms{km s$^{-1}$}
\def\Blos{B$_{\rm los}$}
\def\etal   {{\it et al. }}                     % et al
\def\psec           {$.\negthinspace^{s}$}
\def\pasec          {$.\negthinspace^{\prime\prime}$}
\def\pdeg           {$.\kern-.25em ^{^\circ}$}
\def\degree{\ifmmode{^\circ} \else{$^\circ$}\fi}
\def\ee #1 {\times 10^{#1}}          % \ee p       10^p
\def\ut #1 #2 { \, \textrm{#1}^{#2}} % \ut unit p  unit^p
\def\u #1 { \, \textrm{#1}}          % \u unit     unit
\def\nH {n_\mathrm{H}}
\def\apj{ ApJ}
\def\aap{ A\&A}
\def\mn{MNRAS}
\def\pasp{PASP}
\def\araa{ARAA}
\def\aj{AJ}
\def\apjs{ApJ Supp}

\begin{abstract}
OH(1720 MHz) and methanol masers are now recognized to be excellent
probes of the interactions of supernova remnants with molecular clouds and 
tracers of massive star formation, respectively. 
To better understand the nature of star formation
activity in the central region of the Galaxy, we have used these two classes of masers 
combined with   the IRAC and MIPS data to study
prominent sites of ongoing star formation in the nuclear disk. 
The nuclear disk is characterized by  
massive GMCs with elevated gas temperatures, compared to their  
dust temperatures. We note an 
 association  between methanol masers and a class of mid-infrared  ``green
sources''. These highly embedded YSOs show enhanced 4.5$\mu$m emission 
due to excited molecular lines.
 
The distribution of methanol masers and supernova remnants  suggest 
 a  low efficiency of star 
formation (with the exception of Sgr B2), which we believe is due to  
an enhanced flux of cosmic ray electrons impacting molecular
clouds in the nuclear disk.  
We also highlight the  importance of cosmic rays in  
their ability to heat molecular clouds, and thus increase the gas temperature. 

\end{abstract}

\firstsection % if your document starts with a section,
\section{Introduction}

%\cite[Lee 1971]{Lee71}; \cite{Figer02}).
% NOTE use of \upi in above paragraph and subsequently throughout paper.
% The Greek constant character pi should be upright.
%\cite[Bearman \& Graham (1980, pp.~231--232)]{Bearman80}.
%\section{Green's functions}\label{sec:greenfun}
%\subsection{Construction of equations}

Understanding the processes occurring in the nuclear disk of our own Galaxy is 
interesting not only for insight into our own Milky Way Galaxy, 
but also because it is the 
closest galactic nucleus. This important region of the Galaxy hosts several sources of 
energetic activity, and is the site of massive molecular clouds with pockets of 
 past and present massive star formation. However, the study of the nuclear disk of our 
galaxy has led to a number of paradoxes, puzzles and discrepancies.  
One of the best paradoxical  examples is related 
to the nature of the compact radio source Sgr A* at the dynamical center of the Galaxy. 
Although there is compelling evidence that Sgr A* is a massive black hole (Sch\"odel et 
al.\ 2003; Ghez et al.\ 2003), its X-ray luminosity is orders of magnitude lower than 
expected. Another paradox  is related to the expectation that the formation of stars near 
Sgr~A* should be forbidden due to the extreme tidal forces exerted by Sgr A*; however, many young 
stars are distributed in a disk orbiting Sgr A* (Paumard et al. 2006). Another  
outstanding puzzle is 
the origin of  nonthermal filaments in the Galactic 
nucleus, which has not been understood since their discovery (see Yusef-Zadeh, Hewitt \& Cotton 
2004; Nord et al. 2004).

On a larger scale, the nature of on-going star formation in the nuclear region remains 
unclear. Although there is a high concentration of dense molecular clouds distributed in 
this region, the star formation rate has shown extreme values. On the one hand, the massive 
star forming region Sgr B2 points to the closest example of starburst activity in our 
Galaxy, whereas the quiescent giant molecular cloud (GMC) G0.25+0.01 (Lis \& Carlstrom 
1994; Lis \& Menten 1998) 
appears to contain highly inefficient star formation. 
 This region 
contains some of
 the most spectacular molecular clouds with elevated gas temperatures but 
no obvious source of heating of the clouds. 
In the 
last two decades, studies 
of this region have indicated that the molecular gas is warm, ranging between 75 to 200 K 
(e.g., H\"uttemeister et al. 1993). However, studies based on far-IR observations have 
revealed dust temperature to be 
$<\sim$30K 
(Odenwald \& Fazio 1984; Cox \& Laureijs 
1989;  Pierce-Price et al. 2000). 
This  discrepancy 
between the gas and dust temperatures in the central region of the Galaxy is puzzling.

Using SNR masers  and methanol 
masers, we place constraints on the low efficiency of star formation in the Galactic 
nucleus. We  then argue that the low efficiency of 
star formation and the elevated gas temperature with respect to dust temperature can be 
explained 
by the  enhanced energy density of cosmic rays interacting with molecular 
clouds in the 
nuclear disk.  
The interstellar medium of the nuclear disk 
is distributed in the region with an extent  -1$^0 < l < 2^0$ and -0.5$^0 < b < 0.5^0$
and is characterized by a strong concentration of 
molecular gas in the so-called "Central Molecular Zone''  with a radius of $\sim$200~pc 
(Morris \& Serabyn 1996). 
Physical conditions in this region are extreme, characterized by high velocity 
dispersion 
($\sim$20 km/s), high density ($\sim10^4$ cm$^{-3}$) and high column density 
($\sim10^{23}$ cm$^{-2}$).

\section{Supernova Remnant Masers}

The high number density as well as the  high column density of warm gas 
in the nuclear disk  are similar to the restricted physical conditions 
required to produce  OH(1720 MHz) masers  (Lockett, Gauthier \& Elitzur 1999). 
At the site of C-type shocked gas,  OH abundance is expected to be enhanced sufficiently to 
detect OH(1720 MHz) maser emission throughout the nuclear disk (Wardle 1999).  
With this in mind, we carried  out an unbiased, snapshot   
survey of the Galactic nucleus to search for  SNR masers. When this survey 
is combined with 
other targeted surveys, a total of six SNRs are detected toward roughly the inner 
8$^0\times1^0$ (Yusef-Zadeh et al. 1995, 1996, Frail et al. 1996). However, 
only four   SNR masers
G0.0+0.0 (Sgr A East),  Sgr D (G1.13-0.1), G1.4-0.1 \& G359.1-0.5 are 
detected toward the nuclear disk. 
The only known SNRs with no maser counterparts in the nuclear disk are  G0.9+0.1 
and G0.33+0.01.  Here we discuss SNR masers associated with the SNRs Sgr A East and G359.1-0.5.
% and discuss them. 

\subsection{SNR G0.0+0.0}

The most well known interacting SNR in the nuclear disk is Sgr A East 
(SNR G0.0+0.0) projected toward Sgr A*. The 
kinematics of OH(1720 MHz) masers indicate two populations. 
The first kinematic population has velocities  ranging 
between 30 and 60 \kms \\
distributed around the shell of Sgr A East. Recent sensitive 
observations of OH masers toward Sgr A East have discovered 
 a number of additional maser spots (see the references within Pihlstrom and Sjouwerman  
1996;  Karlsson et al. 2003). 
 These measurements  established that Sgr A East is interacting with molecular gas 
associated with the 50 \kms\ GMC. A number of weak OH(1720 MHz) masers to the 
northeast of Sgr A East  coincide with a large number of shocked 
molecular clumps, as probed by CS (1-0) line observation with NMA (Tsuboi et 
al. 2006), The broad line emission from molecular gas provides additional  support for shocked 
50 \kms\  clouds as a result of the expansion of the SN shock.  The 
line profiles of the CS line emission are asymmetrically blue shifted which implies 
that the far side of the the 50 \kms\ GMC is being shocked by the expansion of the Sgr A East shell  
located behind the 50 \kms\ cloud (Tsuboi et al. 2006).

\begin{figure}
\includegraphics[scale=0.35,angle=0]{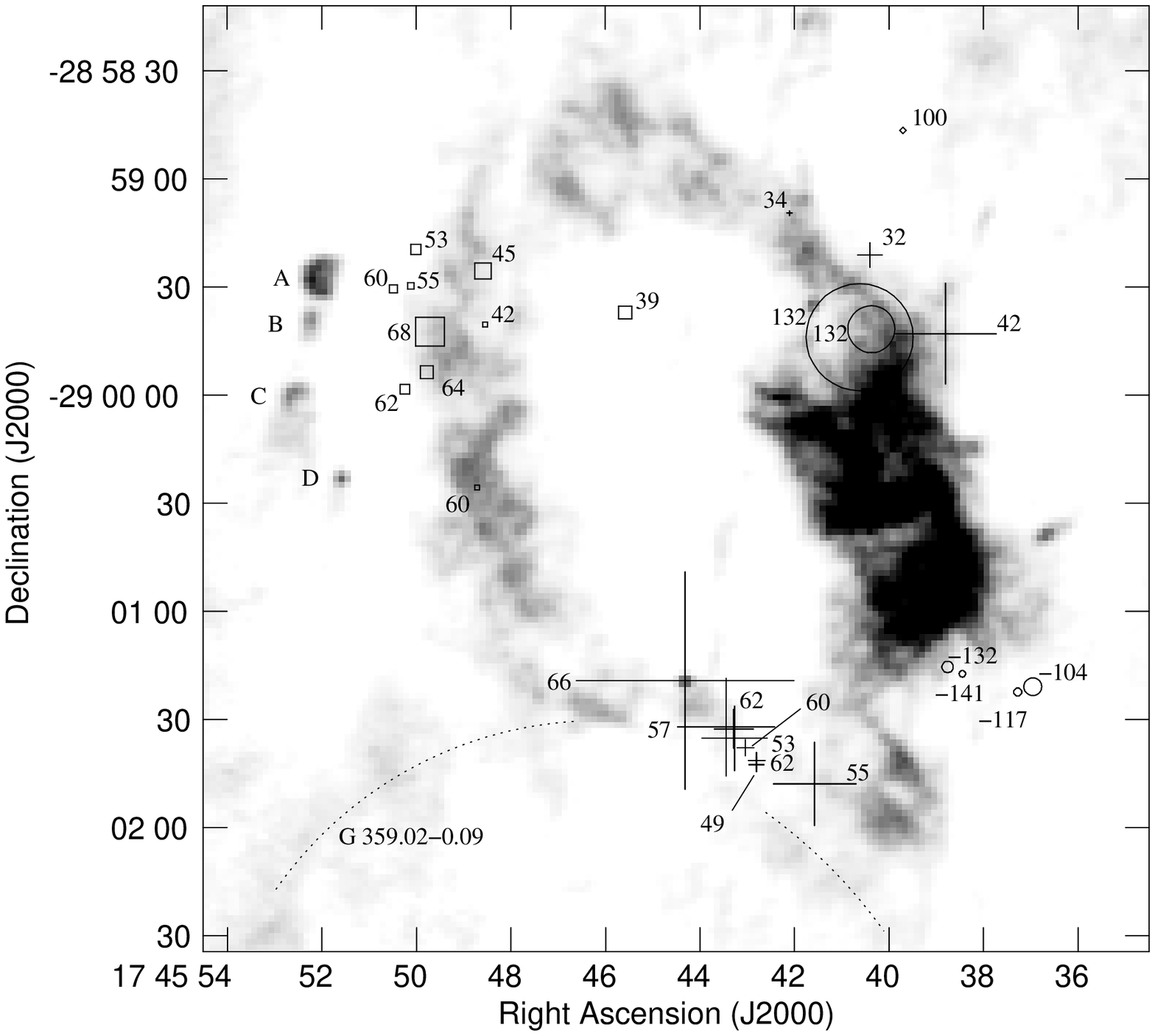}
\includegraphics[scale=0.35,angle=0]{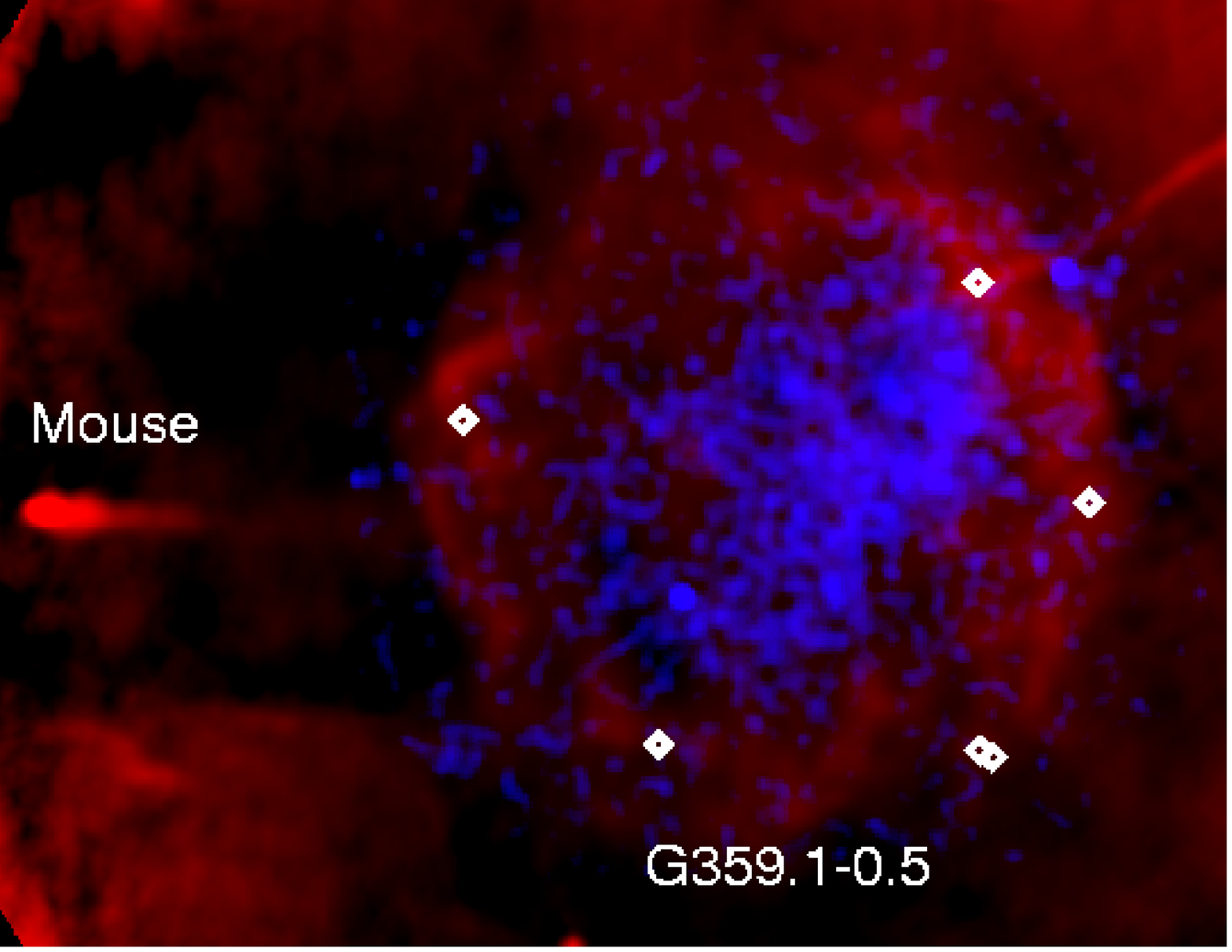}
  \caption{
    (\textit{a})  A grayscale image of the continuum emission from 
                 Sgr A East and West at 20cm. The distribution of 
OH(1720 MHz) maser spots with their V$_{LSR}$ velocities are superimposed on the 
 nonthermal Sgr A East shell 
and the thermal spiral-like structure of Sgr A West (taken from Figure 1 of  Pihlstrom and 
Sjouwerman 2006). 
    (\textit{b})  A 20cm continuum image of G359.1-0.5 (red color)  shows the shell-like 
structure of 
this remnant. The interior of the shell is represented by the distributions of X-ray
 (blue color). The diamond points  represent the positions of maser spots. 
The elongated 
feature to the east of the figure corresponds to radio emission from the pulsar wind nebula known as the 
Mouse.  
}
\label{fig:contour}
\end{figure}

The second kinematic component 
shows  highly blue- and red-shifted  
velocities 
at $\sim\pm$132 \kms (Yusef-Zadeh et al. 2002; Pihlstrom and Sjouwerman 2006). These high 
velocity masers 
are distributed symmetrically at the 
tangent points of the lobes of the  
cicrumnuclear ring orbiting Sgr A*. OH(1720 MHz) masers are
%recognized to be 
excellent indicators of the systemic motion of the molecular
gas that is being shocked. The low velocity dispersion noted between the 
highly blue-  and red-shifted masers places a constraint on  the 
average motion of the enclosed material within the ring of V$_{LSR}\sim0$ \kms. 
  The origin of the masers associated with the circumnuclear ring remains less clear 
than those of Sgr A East. 
On the one hand, it is possible that the expansion of Sgr A East
runs into the circumnuclear ring  and produces the observed OH (1720 MHz) masers. 
This is consistent with the fact that Sgr A East lies behind Sgr A* (Pedlar et al. 1989).
In this picture, Sgr A East lies behind both the 50 \kms molecular cloud and the circumnuclear molecular ring 
and is responsible for production of OH(1720 MHz) masers in both clouds. 
On the other hand, these high velocity maser spots may originate
within the ring by alignment of molecular clumps, resulting in 
amplification along the line of sight (Pihlstrom \& Sjouwerman 2006). Future study of these masers  should 
clarify their relationship to the cicumnuclear molecular ring which is known to be orbiting  
Sgr A*.

\subsection{SNR G359.1-0.5}

%One of the four SNRs showing masers is 
G359.1-0.5  shows  several masers surrounding the 
remnant at a velocity of -5 \kms\ and a low velocity dispersion. 
Such low velocity maser features are unusual if the remnant is interacting 
with molecular gas in the nuclear disk. This low velocity is in contrast to the 
kinematics of a CO shell 
suggested to surround this remnant, which shows velocities ranging between -60 and -150 \kms\ (Uchida et al. 
1992).
 Several more recent studies such as the detections of a bar of H$_2$ emission
at low velocities and  the linear
polarization 
from  the brightest maser spot as well as the relatively low rotation measure of the 
remnant itself 
indicate that G359.1-0.5  is unlikely  be located in the Galactic nucleus. More recently, 
an XMM observation of thermal X-ray gas within the remnant supports a low column density 
to the remnant, consistent with it being a foreground source 
(Heinke et al. 2007, in preparation), as 
described below.  These observations indicate clearly that OH(1720 MHz) maser emission is 
associated with a foreground remnant. This implies that OH(1720 MHz) masers are  
powerful indicators of the systemic motion of the interacting remnant especially in 
the confused region of the nuclear disk. 

The overall
fit to the X-ray  emission from G359.1-0.5 places  a strong constraint on the low-energy 
absorption, 
N$_H=2.6\pm0.2\times 10^{22}$ cm$^{-2}$ (Heinke et al. 2007, in preparation). 
The  pulsar wind nebula known as the Mouse lies only 22 arcminutes from the center of  
the remnant, and pulsar dispersion measurements infer a distance to the Mouse 
between 3 and 5.5 kpc (requiring 
the
Mouse to be within the molecular ring at 3-5 kpc distance, but not beyond the spiral
arm 3 kpc from the Galactic center), giving a nominal value of 5 kpc. 
The N$_H$ to the Mouse is also estimated from X-ray measurements 
to be 2.60$\pm0.09 \times 10^{22}$ cm$^{-2}$ (Mori et al. 2004; Gaensler et al. 2004). 
 Our measured N$_H$ value is consistent with 
 the Mouse's N$_H$, suggesting that G359.1-0.5 is located at a similar 
distance
and is likely to be the origin of the Mouse.  

\subsection{Star Formation Rate from Known SNRs}

If we remove G359.1--0.5 from the list of remnants interacting with molecular gas in 
the 
nuclear disk, then 3/5 or $\sim$60\% of all known remnants in this region are   interacting 
with molecular gas 
in the nuclear disk.  
This is in contrast to the fraction of  interacting remnants 
in  the disk of the Galaxy, 10\%. This implies a uniform distribution of molecular gas 
throughout the nuclear disk. 

Given  the number of known SNRs and assuming a residence time of 2.5$\times 10^4$ yrs, the
SN rate is estimated to be one per 5000 years.  
 This implies 
that the rate of formation 
of massive stars is  
$~10^{-2}$ \msol yr$^{-1}$ (Condon 1992).
This is consistent with the 
far-IR measurements inferring a paucity of massive stars in the nuclear
 disk (Odenwald and Fazio  2004).  This estimate is, however,  quite uncertain since 
the residence time 
of  a remnant  expanding into 
a dense medium is unknown. If the residence time is shorter than the above estimate, 
the star formation rate estimate increases, accordingly. 

\section{Methanol Masers}

It is now well established that methanol masers are signposts  of 
ongoing massive star formation throughout the Galaxy. 
A survey of the inner 2 degrees of the Galactic center 
showed a total of 23  class 
II methanol masers at 6.7 GHz (Caswell 1996). This class of masers is  recognized 
to be 
radiatively pumped, unlike class I methanol masers, known to be collisionally pumped 
(e.g., Menten 1991). 
There is a strong 
correlation between the 6.7 GHz methanol masers and infrared dark clouds as revealed 
in the 8$\mu$m Spitzer observations of the Galactic center (Stolovy et 
al. 2006). 
 The infrared dark clouds are likely progenitors of high mass
stars, and the masers indicate locations where star formation is currently taking place. 
Closer examination of the 
kinematics of the 6.7 GHz methanol masers 
indicates that 11 maser spots arise from Sgr B2, two from Sgr C, and 3-5 from the 
quiescent dust 
ridge G0.25-0.01.  
Given that there is a large reservoir of 
dense molecular gas, up to 10$^8$ \msol, distributed in the nuclear disk, there appears to be 
a relative paucity of methanol masers 
%with a detection sensitivity of 
(above 1 Jy) in the nuclear disk (Caswell 
1996). 
With the exception of Sgr B2, the rarity of bright 6.7 GHz masers suggests low star 
formation activity in the nuclear 
disk. The inner 30 pcs of the Galactic center 
show  no evidence of any class II methanol masers. Remarkably, this inner region is known to 
have 
a large concentration of 
dense giant molecular clouds associated with 
the 20, 40 and -30 \kms\ molecular clouds (e.g., G\"usten et al. 1981). As described below, 
this region is also 
known to be coincident with  strong 
nonthermal emission from Sgr A East, its halo and the 
radio  continuum arc near l$\sim0.2^0$. 

\subsection{``Green'' Sources}

To determine if any of the methanol masers have a stellar counterpart in the mid-IR, we examined 
the color of stars in the nuclear disk. 
One of the most powerful methods  of detecting molecular line emission in mid-IR 
wavelengths using IRAC images is to 
%examine the colors of sources that show
look for excess emission at 
4.5$\mu$m (``green'' in 3-color images combining
3.6, 4.5 and 8 $\mu$m emission).   Using this color selection 
criterion, 
a recent IRAC survey of the Galactic plane has identified a number of interacting supernova remnants 
%with the ISM 
showing  excess emission at 4.5$\mu$m, indicating shocked, excited CO and 
H$_2$ molecular line emission (Reach et al. 2006). 
Also, the ``green nebulosity" has been detected in DR21, a massive
star formation site, in which the excess 4.5$\mu$m color is considered to be due shocked
molecular outflow (Marston et al. 2004; Smith et al. 2006).

\begin{figure}
\includegraphics[scale=0.3,angle=270]{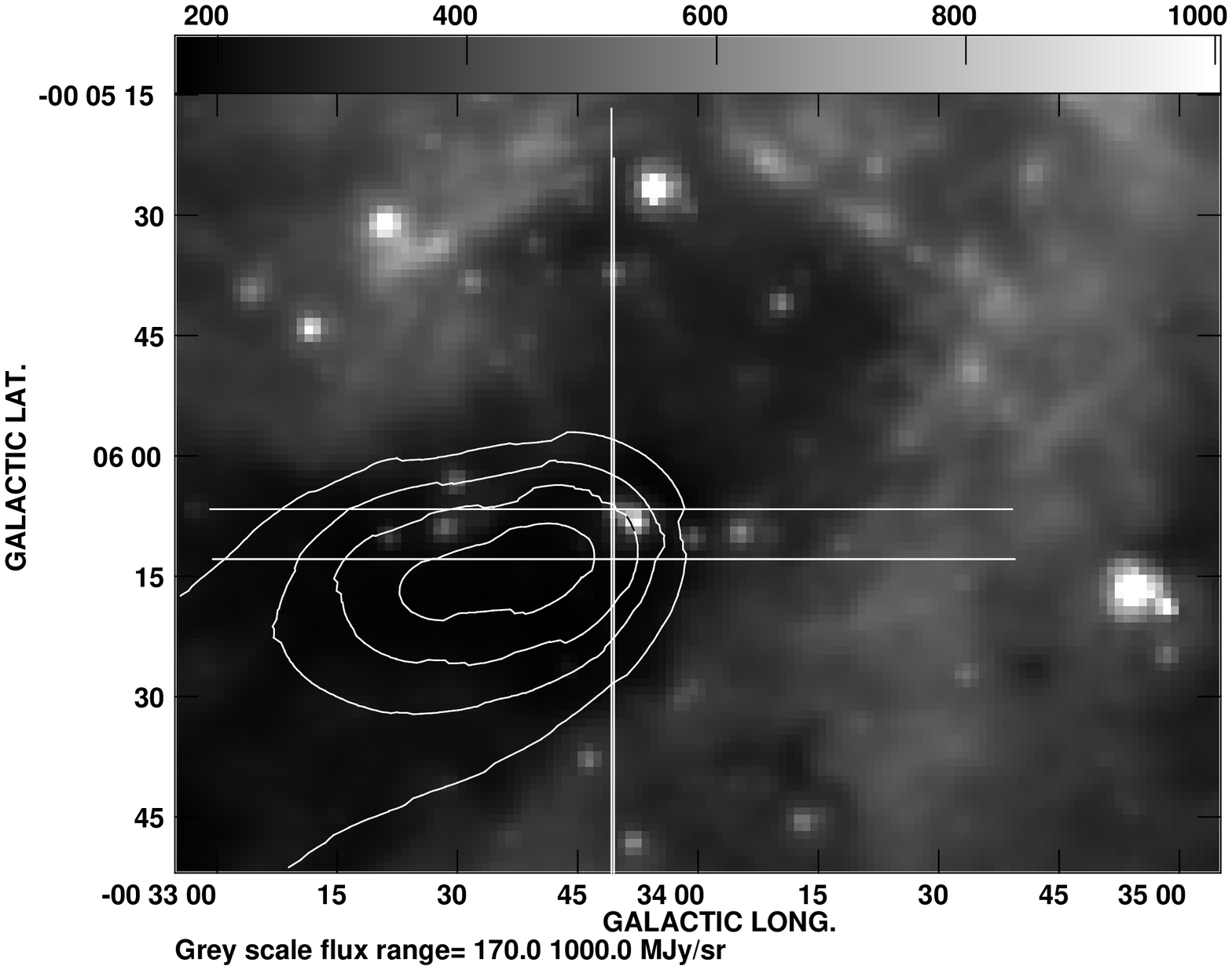}
\includegraphics[scale=0.3, angle=270]{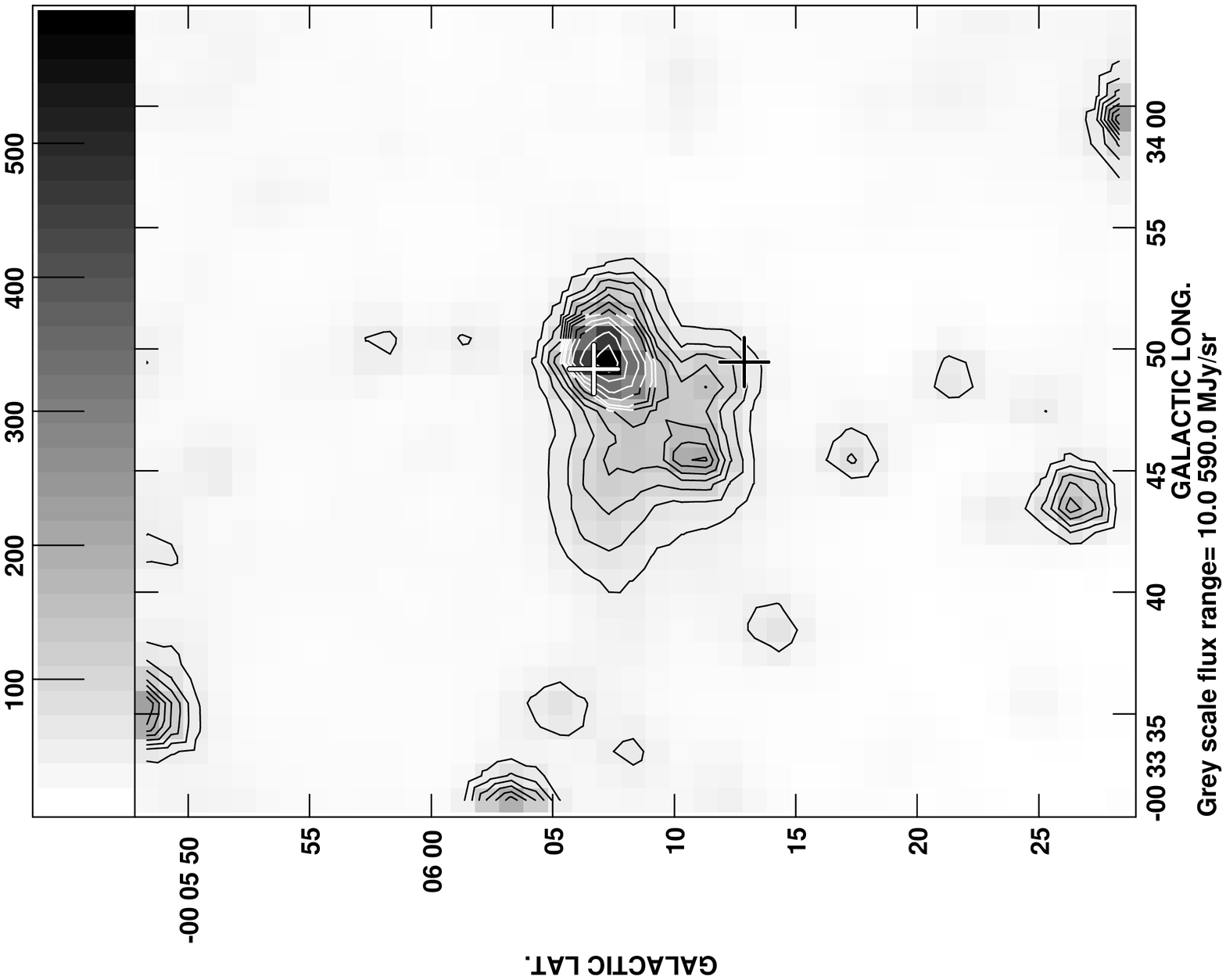}
  \caption{
    (\textit{a})  Contours of 850$\mu$m emission (Pierce-Price et al. 2000) are superimposed
on the 8$\mu$m image of Sgr C. 
    (\textit{b}) Contours of 4.5$\mu$m emission based on Channel 2 of 
IRAC data. The crosses in (a) and (b) represent methanol 
maser distributions associated with Sgr C.
}
\label{fig:contour}
\end{figure}

Using mid-IR data, we have examined the color of stars and dust clouds in star
forming sites throughout the nuclear disk. 
The infrared images of the nuclear disk are generated from the IRAC Galactic center survey
mosaics (Stolovy et al. 2006). 
A quantified indication of excess 4.5 $\mu$m emission is
obtained by constructing the ratio $I(4.5) / [1.2*I(3.6)*I(5.8)]^{(0.5)}$.
This essentially is the ratio of the actual 4.5 $\mu$m intensity to
that determined by a power--law interpolation of the 3.6 and 5.8
$\mu$m intensities. It was found empirically that a slight ($\sim$3\%)
modification to the interpolation coefficients helps emphasize sources
with unusually strong 4.5 $\mu$m emission. In a map of this ratio
across the entire Galactic center survey, most point sources exhibit a uniform ratio
that is not very different from the background. In regions of high extinction
(e.g. IR dark clouds, and Sgr  B2), this ratio rises because the reddening
reduces the 3.6 $\mu$m intensity. However, 
there were  of order 100 sources, located in regions of both high and low extinction, which exhibit ratios distinctly
higher than those of  other sources in their vicinity (within several arc minutes).
These include 33 sources that
were visually selected
as "green" sources in 3-color images. Yet the ratio also clearly
identifies a number of interesting
sources that do not stand out in the 3-color images, such as 3 planetary
nebulae. There
are thirty green sources suggesting YSOs embedded within infrared dark
clouds.
A correlation of methanol masers in the nuclear disk with the  green sources
has led to the surprising result that
about 1/3 of green sources have maser counterparts, 1/3  have no maser counterparts,
%has not been observed previously 
 and the rest have  not yet been observed. 
Given the limited sensitivity of 1 Jy in the maser survey by Caswell (1996), 
it is possible that many of the green sources have weak maser counterparts or that they 
signify early sites of  massive star forming regions.

Although  the  $\sim2''$ spatial resolution of
IRAC images being different than the 2$''\times4''$  resolution of  methanol maser
observations (Caswell 1996), we have mapped the data with 1$''$ pixels. 
The green sources are generally seen to be extended on a scale $\sim10''$ and 
the maser  sources 
fall within 
the extent of the 4.5$\mu$m emission. 
 An example in which the relationship between a green source and a methanol maser   
can be seen is in the Sgr C complex region. This complex is known  to  
consist of an evolved, extended  HII region and a prominent nonthermal radio filament 
running perpendicular to 
the Galactic plane (Liszt \& Spiker 1995; Yusef-Zadeh, Hewitt \& Cotton 2004). At the core 
of the HII region,  an elongated 
infrared dark cloud is 
noted in which two maser sources and a green source are embedded. Figure 2a shows 
contours of submm emission at 850$\mu$m (Pierce-Price et al. 2000) superposed on an IRAC image of this 
region at  
8$\mu$m. 4.5$\mu$m contours of the bright source which shows prominently in a three-color 
image of Sgr C  are shown in Figure 2b. The crosses show the positions of 
methanol maser emission at 6.7 GHz. As pointed out earlier both these maser sources
could be associated with the green source in Sgr C. We have also 
mapped 44-GHz methanol maser emission from the green source using Haystack Observatory. 
This transition of methanol maser emission, which is known to be collisionally pumped,  
is identified on a scale of $1'$ oriented along the extent of the  major axis of the 
green  source.  This is also consistent with a picture in which the green source is 
associated with the maser emission source in Sgr C  (a more detailed account of this 
source will be given elsewhere). 

\begin{figure}
\includegraphics[scale=0.35,angle=0]{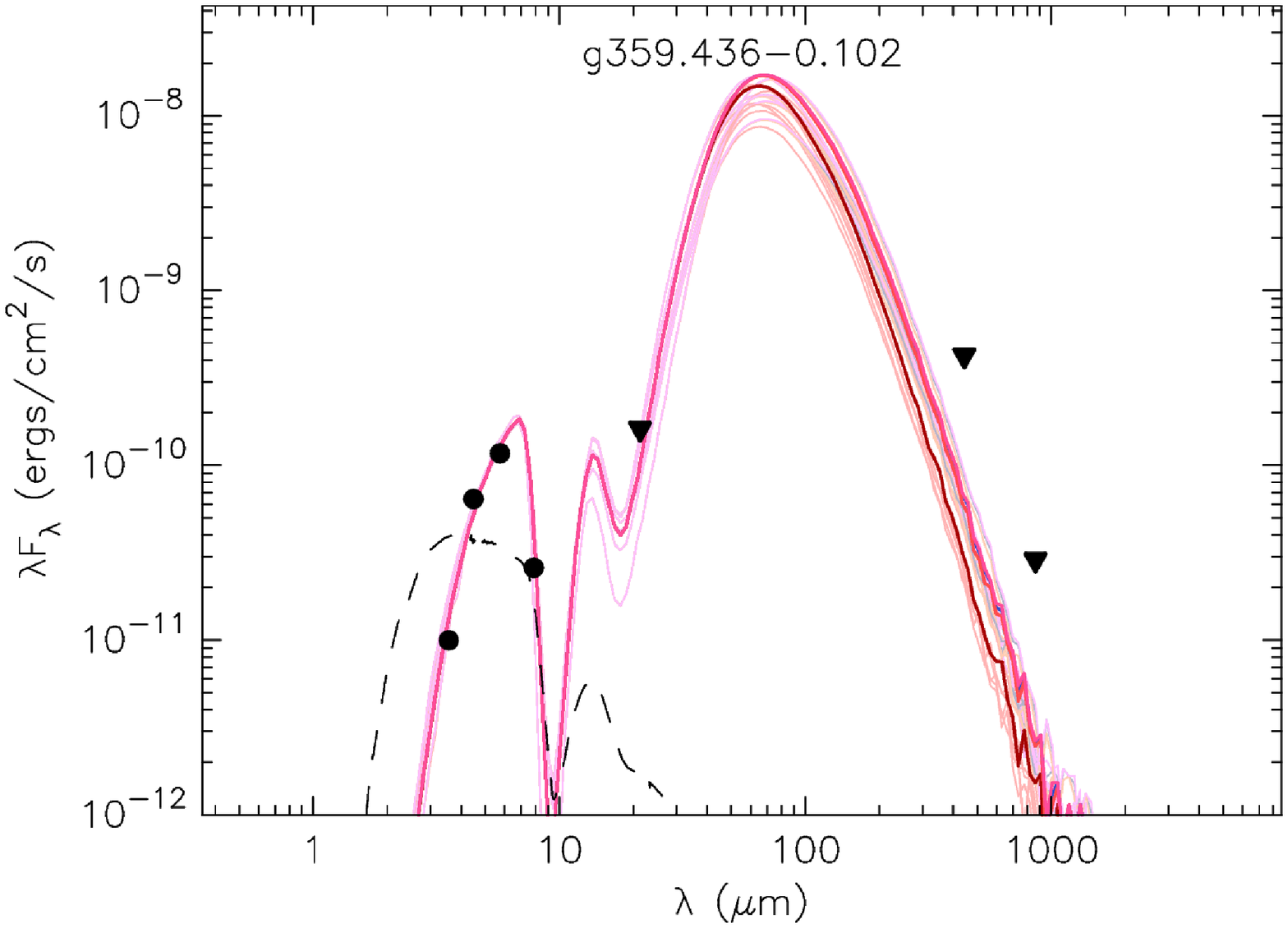}
\includegraphics[scale=0.6, angle=0]{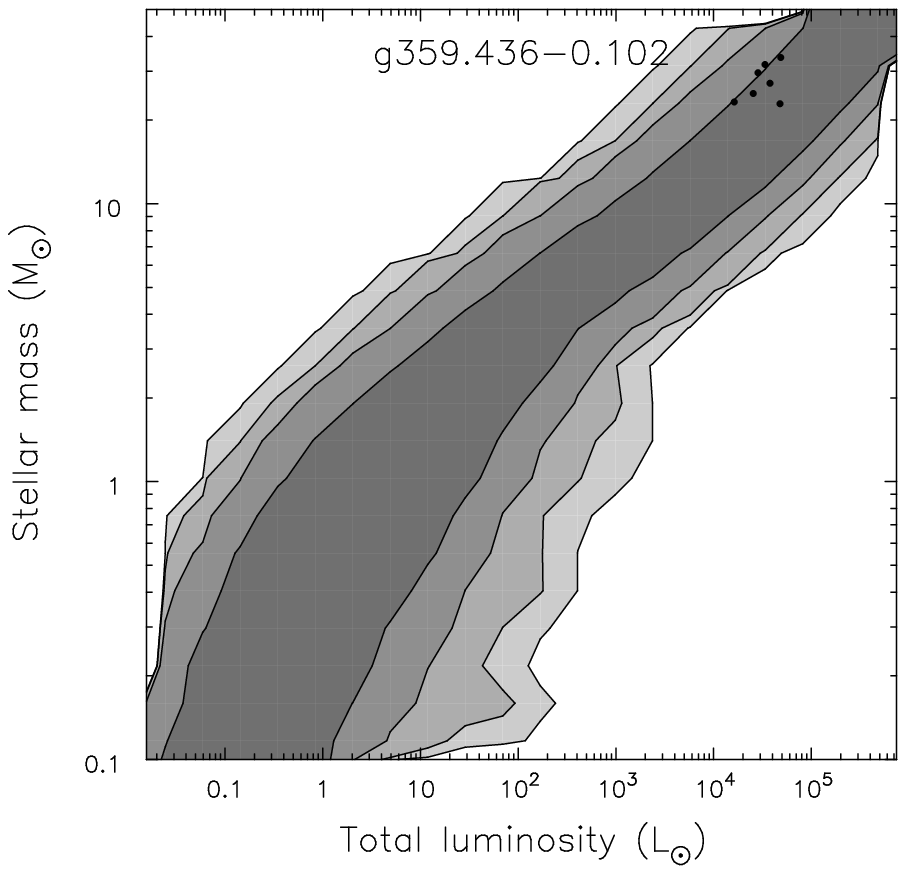}
  \caption{
    (\textit{a}) A preliminary  SED of the northern maser G359.436-0.102 in Sgr C and the 
fitted model
with different aperture sizes.   
    (\textit{b})  The grey scale shows distribution of stellar mass and
luminosity from a grid of 20,000 YSO radiative transfer models.  The black
dots show the mass and luminosities of the well-fit SEDs (those in Figure
3a).
}
\label{fig:contour}
\end{figure}

We have modeled the SED of the green source G359.436-0.102
 in Sgr C and have estimated the
mass and luminosity of the central
protostar coincident with the norther maser source by fitting the SED from a
large grid of 2-D YSO models (Robitaille et al. 2006, 2007).  The models
include a disk and an accreting envelope with bipolar cavities.
Figure 3a shows the best fitted models to the data using IRAC, MIPS and submm data 
accounting  for 30 magnitudes of visual extinction.   
Due to the lower resolution of submm data, we show upper limits for the 
peak flux densities at 850 and 450 $\mu$m (Pierce-Price et al. 2000) at the position of the 
green source. 
Different 
curves of the model fits
correspond to the variation of the observed  apertures. 
These model fits  all seem to be consistent with the SED of the source.
Figure 3b shows the derived mass and luminosity of the green source 
based on the well-fit SED  models.
This fit of the SED  
% data points to the green source
 is  consistent with the identification of  
the green source  with a massive protostar,
associated with the methanol masers.
% and that the green source and the methanol masers are associated with each other. 
%The    correlation of green sources and methanol masers
The derived parameters of protostars using green sources with methanol maser
counterparts
should be useful indicators  of the evolutionary phase of these sources. 
In addition,
further study of the green/maser sources may also
  address  whether linear distribution of  maser
spots come from circumstellar disks or outflows (De Buizer et al. 2005). Furthermore, green
sources found based on  IRAC  observations  can provide an extensive target list for future
methanol maser studies.

\section{The Role of Cosmic Rays}
We believe that the  low on-going star formation activity 
inferred from  the estimate of the supernova rate and the rarity of methanol masers 
can be explained by enhanced cosmic ray flux in the nuclear disk.
High cosmic-ray fluxes in molecular clouds affect star formation by
heating the gas and increasing its ionization fraction (Yusef-Zadeh, Wardle \& Roy 2007a).  
Higher cloud
temperatures increase the Jeans mass, potentially changing the IMF,
while high ionization increases magnetic coupling to the cloud
material, reducing ambipolar diffusion and increasing the time
taken for gravitationally unstable cores to contract to the point that
they overwhelm their magnetic support. 

Several studies have recently indicated the importance of cosmic rays
in the nuclear disk. One is from  the 
detection of low frequency 74 MHz radio emission from the  
central region  of the Galaxy,  indicating 
 enhanced cosmic rays with cosmic ray electron density of $\sim$7 eV cm$^{-3}$
(LaRosa et al.  2005).  The other is 
strong H$_3^+$ absorption lines detected along several lines of sight towards the
Galactic center (Oka et al.  2005). These absorption lines are 
much stronger than those  seen in the Galactic disk. These measurements infer 
that  the ionization rate in the inner 30pc of the Galaxy 
 is more  two orders of magnitude higher than 
that that in the 
Galactic disk. The high ionization rate is more likely to be 
due to high cosmic ray flux as revealed by high synchrotron emissivity in raido wavelengths 
than 
due to diffuse hot X-ray emitting gas.  
Lastly,  the large scale distribution of K$\alpha$ Iron   emission line at 6.4 keV 
associated with a number of cloud complexes in the nuclear disk has been correlated 
with the distribution of nonthermal radio emission from this region (Yusef-Zadeh et al. 
2007b). The energy density of the
cosmic rays required to explain the 6.4 keV X-ray emission from the inner
2$^0\times0.5^0$ of the molecular nuclear disk 
ranges between 20 and 10$^3$
eV cm$^{-3}$. 

% which is considerably higher than that in the local ISM of 0.2 eV cm$^{-3}$ (Webber 
%1998). 

%From (\ref{Helm}), the fundamental wave solutions to (\ref{inteqpt}) and

%\section{Conclusions}\label{sec:concl}

\begin{acknowledgments}
%the solution procedure used in \S\,\ref{sec:concl}. A.\,N.\,O. is supported
%by SERC under grant number GR/F/12345.
We thank Y. Pihlstrom for providing us with Figure 1a. C.O.H. is supported
by the Lindheimer Postdoctoral Fellowship. 
\end{acknowledgments}

\end{document}